\documentclass[final]{aa}

  \usepackage{amscd}
  \usepackage{amsmath}
  \usepackage{amssymb}
  \usepackage[T1]{fontenc}
  \usepackage[varg]{txfonts}
  \usepackage{graphicx}
  \usepackage{natbib}
  \usepackage{verbatim}
  \usepackage{array}

\usepackage{aas_macros}

\newcommand{\mathbi}[1]{\mbox{\boldmath$#1$}}

\begin{document}

\title{A Model for Multidimensional Delayed Detonations in SN Ia Explosions}
\titlerunning{Multidimensional Delayed Detonations}

\author{I.\ Golombek \and J.C.\ Niemeyer}
\institute{Lehrstuhl für Astronomie, Universität Würzburg, Am Hubland, D-97074
  Würzburg, Germany}

\abstract{
We show that a flame tracking/capturing scheme originally developed for
deflagration fronts can be used to model thermonuclear
detonations in multidimensional explosion simulations of type Ia
supernovae. After testing the accuracy of the front model, we present
a set of two-dimensional simulations of delayed detonations with a
physically motivated off-center deflagration-detonation-transition
point. Furthermore, we demonstrate the ability of the front model to
reproduce the full range of possible interactions of the detonation
with clumps of burned material. This feature is crucial for assessing
the viability of the delayed detonation scenario.

}

\maketitle

\section{Introduction}
The widely accepted standard model for type Ia supernovae (SNe Ia) is
the thermonuclear explosion of a C+O white dwarf that has reached the
Chandrasekhar mass by means of mass accretion from a binary
companion  \citep[e.g.][and references therein]{HN00}. This scenario
has recently received new support from the tentative discovery of the
companion star of Tycho's supernova by  \citet{Rea04}.

While this idea has been around since the original work by
\citet{HF60} and countless CPU cycles have been spent in search of a
realistic model, some crucial aspects of the explosion physics are
still not understood. Apart
from the scientific challenge of reproducing the inner workings of these
fascinating astronomical events of which we already know so many details 
\citep[e.g.][]{Mea04,Bea04b}, their potential as cosmological distance
indicators has provided us with yet another strong motivation to
construct predictive, self-consistent explosion models. Two subjects
take center stage in most current debates: (i) the conditions under
which the dynamical phase of the explosion ignites and (ii) the
thermonuclear flame forms \citep{WWK04}, and the questions surrounding
the possibility of a deflagration-detonation-transition (DDT)
\citep{NW97,KOW97} during the late phase of the explosion, giving rise
to a ``delayed detonation'' \citep{WW94a,K91a,Yea92}. The modeling and
consequences of the latter are the topics of this paper.

Detection of intermediate mass elements in the early spectra of
SNe Ia is usually interpreted in terms of an initial phase of subsonic
thermonuclear burning (deflagration) that allows the star to expand to lower
densities. The thin deflagration front is hydrodynamically unstable,
most importantly as a result of the Raleigh-Taylor (RT) instability,
so it quickly becomes fully turbulent. Different models for the
unresolved turbulent flame structure have been employed in
multidimensional simulations of exploding white dwarfs
\citep{NH95a,K95,RHN99}; however, the global properties of the explosion
appear to be robust with respect to the particular choice of
subgrid-scale model \citep{RHN02b,GKOe03}. 

On the other hand, interpretations of the simulations with regard to
whether or not a DDT must occur differ considerably. Pure deflagration models
with the currently achievable grid resolution produce too little
$^{56}$Ni when compared with one-dimensional models that successfully fit the
observational data. Moreover, they generically leave behind
unburnt material near the stellar core, whose amount is strongly
constrained by the lack of carbon features in late SN Ia spectra. It
has been claimed that this is a problem intrinsic to all turbulent
deflagration models which can only be solved by a delayed detonation
that burns up all of the C+O left in clumps of varying sizes near the
core \citep{GKO04}.

Owing to the fine-tuning required to achieve a DDT in unconfined media
(\citealt{N99}; strengthened by the recent results of \citealt{Bea04c}) and to
the promising trend in highly resolved turbulent 
deflagration models for burning progressively more material in the core
\citep{NRTH03}, we believe that the jury is still out on the question of
delayed detonations. Nevertheless, we will demonstrate in this work
that the level-set method that has been succuessfully employed to model
deflagration fronts in SN Ia simulations \citep{RHNe99} can be
modified in a straightforward way to represent unresolved
detonations. Its main advantage is its complete control over the front
velocity at each point, allowing us to mimic the interaction of the
detonation with unresolved and resolved clumps of ashes in a realistic
manner. One of our main conclusions is the sensitivity of the results
with respect to the way these interactions are implemented, supporting our
claim that an accurate detonation model is necessary.

In previous attempts to simulate delayed detonations in multiple
dimensions \citep{AL94b,GKO04}, the detonation front was unresolved and no
model for the propagation speed was employed. While it is well-known
that the velocity of planar detonations is robust with regard to
numerical resolution, the same is not true in the presence of
obstacles or clumps of ash (closely related are 
consequences of the cellular front structure, see
\citealt{Tea00}). In fact, fully resolved calculations show that C+O
detonations can hardly pass through burned material at all,
whereas this behavior is easily hidden by coarse resolution
\citep{M05}. Here, our approach is to compare the two limiting cases
of (i) a detonation that passes through burnt material as a pure shock
wave at the speed of sound and re-ignites on the other side and (ii) one
that stalls immediately when it hits a clump of ash and needs to wrap
around it to reach the other side. 

In all of our simulations, we let the code decide where and when the
DDT should take place. The criterion for a DDT follows from the
comparison of the thermal flame thickness with the Gibson length of the
turbulent flame brush \citep{NW97,NK97a}. As expected, we find that
all DDTs occur near the outermost parts of the turbulent
deflagration front, generally taking place earlier in off-center
ignition models than in centrally ignited ones. It follows immediately
that in order to burn the clumps of C+O near the core the detonation
needs to propagate back into the star, having to pass through a foam of
fuel and ashes along the way. This is the reason why an accurate
treatment of the detonation-clump interactions is indispensable.

In what follows we give a short description of the flame algorithm
employed to model both the deflagration, as well as the detonation
front. In Section \ref{sec:test}, we discuss the results of a series of
test calculations determining the suitability of our approach to
represent supersonic burning fronts. In Section \ref{sec:results}, we
present our DDT model and discuss the results and
implications of our numerical study. Concluding remarks are made in
Section \ref{sec:conclusions}.  
 
\section{Level set deflagration and detonation model}

The numerical code applied in this work is based on 
Reinecke's version of PROMETHEUS as described in \citet{RHN99,RHNe99}. In
particular, we employ the level set capturing/tracking scheme \citep{S96}
used in these 
references for turbulent deflagration fronts to model an additional
delayed detonation. In this 
approach, the thermonuclear burning front is associated with the zero level
set of a linear distance function $G(\mathbi{x},t)$. It thus represents a 
$n-1$-dimensional moving hypersurface in an $n$-dimensional
simulation. For a similar approach to model deflagrations and detonations, see
\citet{FAX99}. 

The front propagation is given by the temporal evolution of $G$,
\begin{equation}
\label{equ:g}
\frac{\partial G}{\partial
t}=(\mathbf{v}_{\mathrm{u}}\cdot\mathbf{n}+s)|\nabla G|\,,
\end{equation}
depending on the fluid motion $\mathbf{v}_{\mathrm{u}}$ of the
unburnt state and the propagation speed $s$ relative to it. The choice of $s$
reflects the kind of combustion front, whether deflagration (subsonic) or detonation
(supersonic), and is provided as an external parameter as described in the
next paragraph. Details about the level set function and its implementation
can be 
found in \citet{RHNe99}. For the modeling of both deflagration and detonation
fronts we used the cell averaged fluid velocity $\mathbf{\bar{v}}$ instead of
$\mathbf{v}_{\mathrm{u}}$. Owing to this so-called passive implementation, as
opposed to the full implementation, cf.\ \citet{SMK97,RHNe99}, the
transition between fuel and ashes is smeared out over about three grid cells.

As in previous simulations \citep{RHN99,RHN02b}, the speed of the
deflagration front is defined 
as the upper limit of the laminar and turbulent burning velocities 
in the respective grid cell,
$s_{\rm def}=\mathrm{max}(s_{\mathrm{lam}},s_{\mathrm{tur}})$. 
Here, $s_{\mathrm{lam}}$ is a known function of composition, density,
and temperature \citep{TW92}, while $s_{\mathrm{tur}}$ is
calculated by an additional subgrid-scale model taking fluid dynamics into 
account on scales smaller than the computational grid size
\citep{NH95a}. The detonation velocities, $s_{\rm det}$, are taken from
\citet{S99} as a function of the unburned fuel density. 

Behind the front all material is immediately transformed into the
reaction products, whereas inside the mixed cells we determine the volume
fraction  
occupied by the burnt material and adapt the mass fraction of
ashes correspondingly. For densities
$\rho>5.25\times10^7\,\mathrm{g/cm^3}$, the fuel is converted into
$^{56}\mathrm{Ni}$, releasing a fusion energy of
$7.86\times10^{17}\,\mathrm{erg/g}$. At lower densities, the burning
process only produces intermediate mass elements represented in our
simulations by $^{24}\mathrm{Mg}$ and an energy
release of $4.18\times10^{17}\,\mathrm{erg/g}$. Below 
$\rho=10^7\,\mathrm{g/cm^3}$, no change of composition takes place.

\section{Detonation test calculations}
\label{sec:test}

Since the front algorithm had so far only been applied for modeling
deflagrations, its suitability for describing detonations
had to be verified. Several one- and two-dimensional test
calculations were carried out to investigate the influence of external
factors like fuel density, grid resolution and time step on the front
propagation. 

In all simulations, the unburnt material was characterized by a uniform
composition with equal mass fractions of $^{12}\mathrm{C}$ and
$^{16}\mathrm{O}$ and an initial temperature of
$T=10^8\,\mathrm{K}$. Unless otherwise stated, all fronts were set 
to propagate in a positive $x$-direction (and a positive $y$-direction
in the 2D case) on a Cartesian grid with a cell size of
$\Delta=10^6\,\mathrm{cm}$. 

\subsection{One-dimensional fronts}

\begin{figure*}[t]
\centerline{
\includegraphics[width = \linewidth]
  {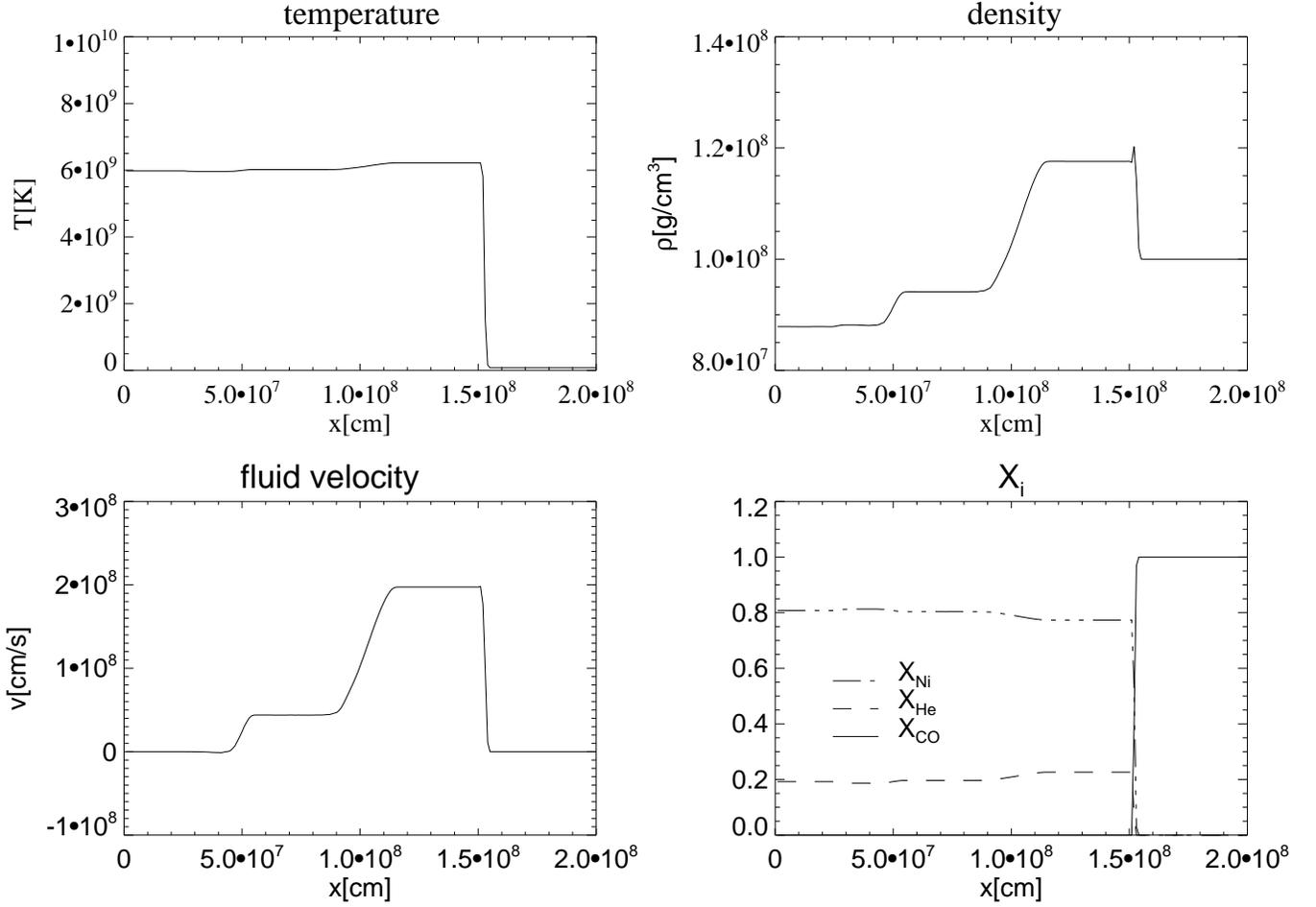}
}
\caption{Temperature, density, fluid velocity, and chemical composition
of a planar detonation front after 0.1 s. The spike in the density
profile behind the shock is the result of a small numerical oscillation.}
\label{fig:detonation01s}
\end{figure*} 

We investigated planar fronts propagating on a 
$200\times 4$ grid with reflecting boundaries at the left, top, and
bottom edges of the computational domain and an outflow boundary to
the right. Figure \ref{fig:detonation01s} shows the profiles of 
temperature, density, fluid velocity, and chemical composition of a 
detonation front 0.1 s after its initialization. 

In order to investigate the ability of the front model to reproduce
a given propagation velocity $s_{\rm det}$, we compared the time dependent 
$x$-positions of the detonation front with its theoretically predicted 
behavior for three different initial densities $\rho=10^7, 10^8$, and
$10^9\,\mathrm{g/cm^3}$. To determine the position of the front, the 
values of the level set function in the direct neighborhood of the
front were interpolated. 

Additionally, to verify the
obtained $x$-values, we directly calculated the propagation 
velocity $\tilde s_{\rm det} $ by averaging the mass fraction of ashes per
cell over  
all 200 grid zones in the $x$-direction in every time step, multiplying 
with the physical grid length and dividing by time:
\begin{equation}
\tilde s_{\rm det} = 200\,X_{\mathrm{ash}}(t) \frac{\Delta}{t}\,.
\end{equation}
Here, $X_{\mathrm{ash}}(t)$ denotes the averaged time dependent mass
fraction of ashes. Both methods gave virtually identical results.
For all three densities, the measured numerical velocity of the detonation
front $s_{\rm det}$  exceeded the provided one, $\tilde s_{\rm det}$, by
approximately 10 \%. As shown by further numerical 
tests, this behavior is affected neither by a change of the grid 
resolution nor by a considerable refinement of the time step, so that the error
can be attributed to the inaccuracies in the passive implementation of the
front model.
 
Reconsidering the velocity profile in Fig.\ \ref{fig:detonation01s}, the 
reason for the excessive propagation speed becomes clear. In
our simulation, the fuel is at rest, while the ashes behind the front 
proceed with $v_{\mathrm{b}}\approx 2 \times 10^8$ cm s$^{-1}$. 
Averaging the two velocities results in a non-zero fluid velocity and
therefore causes an error in Eq.\ \ref{equ:g} when replacing
$\mathbf{v}_{\mathrm{u}}$ by $\mathbf{\bar{v}}$.

Since in these numerical tests the fuel had initially been at rest, the 
results shown above are only valid in the frame of reference of a
non-moving gas. In order to test for a possible grid dependence of the front
propagation scheme, the system of detonation front and gas was
Galilei-transformed in a further simulation. Here, the C+O 
gas flowed uniformly in the negative $x$-direction with a velocity 
$v_{x}=-1.16\times10^9$ cm s$^{-1}$, and the reflecting
boundary condition at $x=0$ was changed to outflow. The front was ignited at
the center of the grid. Since $|v_{x}|$ was chosen to be equal to $s_{\rm
  det}$  at the density $\rho=10^8$ g cm$^{-3}$, the front should remain
at rest in the grid 
frame on its right hand side while propagating with double velocity to the
left.  

\begin{figure*}
\centerline{
\includegraphics[width = \linewidth]
  {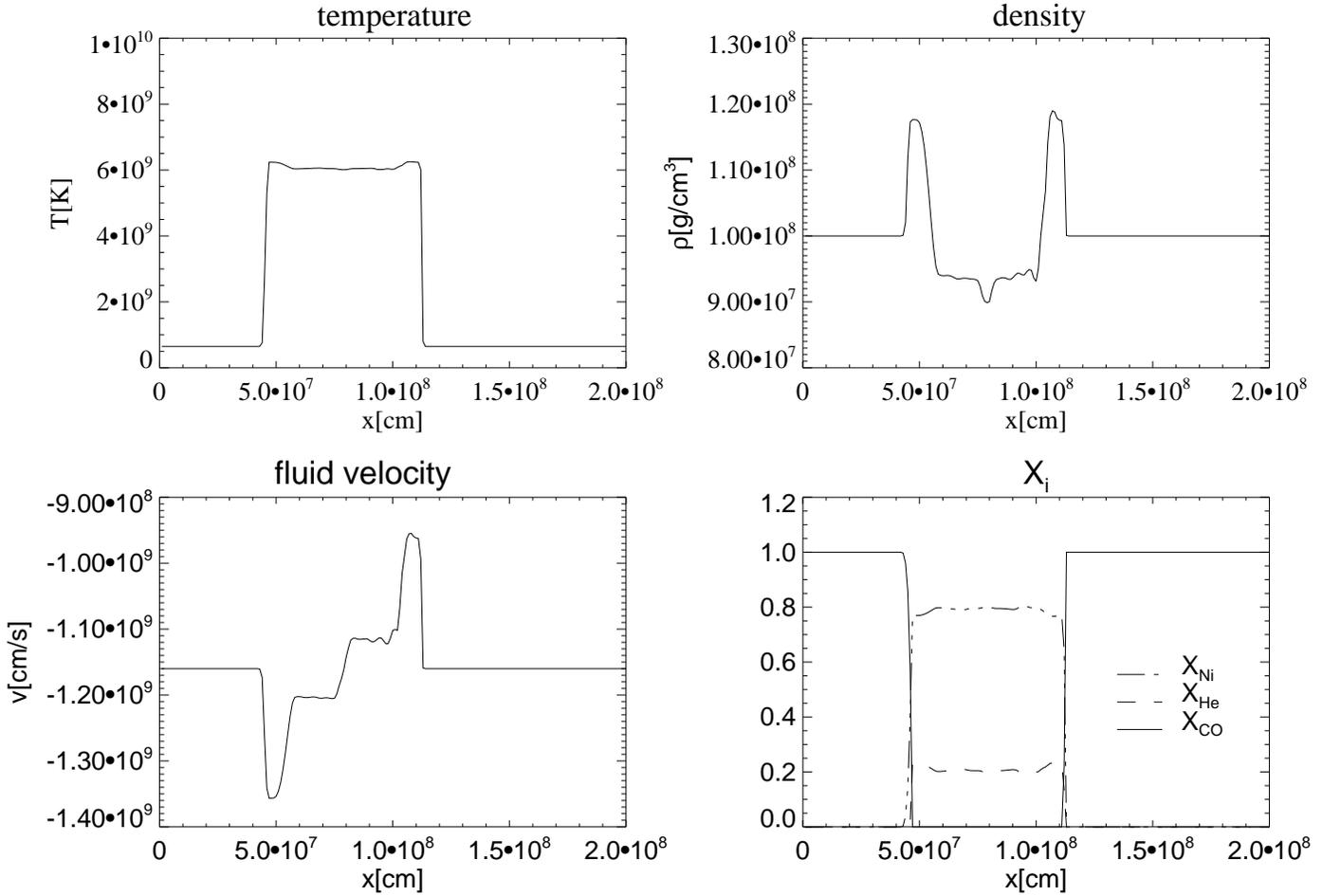}
}
\caption{Detonation profile $0.018\,\mathrm{s}$ after
initialization in a C+O gas of velocity 
$v_{x}=-1.16\times10^{9}$ cm s$^{-1}$.}
\label{fig:galilei}
\end{figure*}

The profile of the front $0.018\,\mathrm{s}$ after its initialization
can be seen in Fig.\ \ref{fig:galilei}. 
The measured deviation ($\approx 7$ \%) is equal in both directions, showing
that the front model is insensitive to Galilei transformations. This fact
allows a simple correction of $s_{\rm det}$ to account for the numerical error.

\subsection{Two-dimensional fronts}

\begin{figure}[t]
\centerline{
\includegraphics[width = \linewidth]
  {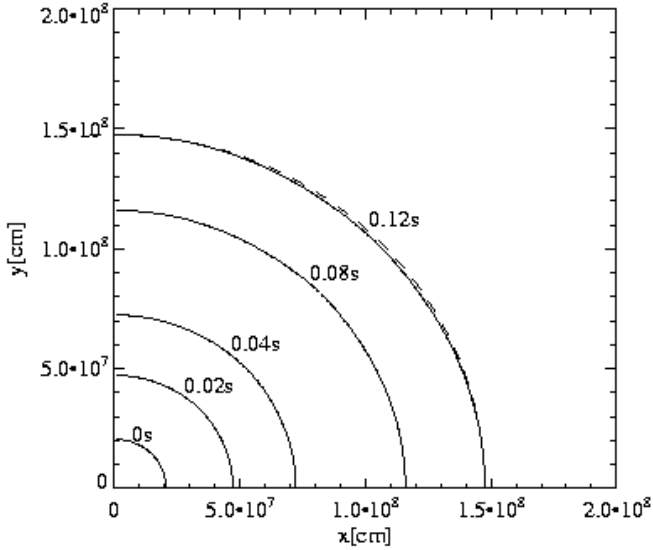}
}
\caption{Front geometry of a centrally initialized circular front
propagating outwards. The dashed line represents an exact circle.}
\label{fig:kugelfront}
\end{figure}

For the investigation of fronts in two dimensions, the grid was expanded
from 4 to 200 cells in the $y$-direction.
Calculations with circular detonations showed that the front kept 
its shape throughout the simulation time, differing  
insignificantly from the exact circle geometry (Fig.\ 
\ref{fig:kugelfront}). As in the case of planar detonations, the 
small deviations of the front velocity from its theoretical value can
be explained by the fact that the front is not advected with the speed
of the unburnt matter in Eq.\ (\ref{equ:g}) but by the average speed 
in the burning cells.

Comparing our findings to the results of \citet{RHNe99}, one 
can see that the front algorithm causes errors like those 
for both kinds of burning fronts, with the difference 
that the speed is too low in the case of deflagrations while
being too high in the case of detonations. The reason 
for this opposite behavior is that in the earlier work, all material 
behind the front is at rest while the fuel expands at high velocities. 
Underestimation of the propagation speed  is therefore a 
consequence of averaging the fluid velocities in the cells cut by the 
front. In contrast to our previous expectations, the higher density jump across
the  front does not result in a larger inaccuracy for detonations. 

\section{Two-dimensional simulations of delayed detonations}
\label{sec:results}

The supernova simulations were carried out in cylindrical 
coordinates on a Cartesian grid in the $r$-$z$-plane. The center of
the white dwarf was placed into the origin of a $256\times256$ mesh
with equally spaced cells of a side width $\Delta=10^6\,\mathrm{cm}$.
Thus, we modeled one quadrant of the star, assuming mirror symmetry 
along the $r=0$ and $z=0$ axes. To take the expansion of the star into 
account during the explosion and to avoid violating 
of mass conservation due to an outflow of stellar material over the 
grid borders, the size of the outermost grid zones grew exponentially 
beginning with the 222nd zone. 

The white dwarf, constructed in hydrostatic 
equilibrium for a realistic equation of state, had a central density 
of $2\times10^9$ g cm$^{-3}$, a radius of about 
$1.9\times10^8\,\mathrm{cm}$, and a mass of $2.8\times10^{33}\,\mathrm{g}$.
The initial mass fractions of C and O were chosen to be
$X_{\mathrm{C,O}}=0.5$, and the total binding energy turned out to be
$-5.2\times10^{50}\,\mathrm{erg}$.

The deflagration phase of the models Z1, Z3, B1, and B5 was initialized and
evolved identically to \citet{RHN99}.
For modeling the delayed detonation, a second level set function
(LSet2) was implemented in addition to the first one (LSet1) 
describing the deflagration front. In order not to disturb the flame 
propagation during the deflagration phase, the value of LSet2 was 
set to a negative value. 

A DDT was assumed to occur as soon as the transition from flamelet to
distributed burning regime took place for the first time during the
explosion \citep{NW97,NK97a}. In order to identify the corresponding time and
grid location, the thermal flame width $\delta$ was compared to the Gibson 
length, defined by
$l_{\mathrm{Gibs}}=\Delta\left(s^2_{\mathrm{lam}}/2q\right)^{3/2}$, 
in each time step. The values of the flame width as a function of 
density were taken from \citet{TW92}. A detonation 
was initialized as a circular front with a radius of
$10^6\,\mathrm{cm}$ at the point where the condition
$l_{\mathrm{Gibs}}=\delta$ was first satisfied.

So far, the question whether a thermonuclear detonation
front can propagate through burnt stellar material has not received much
attention. It is conceivable that the pressure wave is strong enough to
ignite the carbon again when it re-enters a region of fuel. However, fully
resolved simulations suggest that this is not the case even for relatively
small clumps of ash \citep{M05}.  
Depending on the behavior of the detonation front, isolated bubbles 
of fuel may be left inside the stellar core even after a delayed detonation
has taken place, in disagreement with spectral observations in the nebular
phase. 

In order to test the capability of our detonation model to cover the whole
range of expected front behavior, we investigated the limiting cases of a
detonation that first crosses the burnt material as a shock wave 
with sound velocity and then re-ignites on the other side (Case a), and a
detonation that dies immediately upon running into ashes (Case b). The
latter was modeled by setting $s_{\rm det} = 0$ in burnt regions.

\begin{figure*}
\centerline{
\includegraphics[width = \linewidth]
  {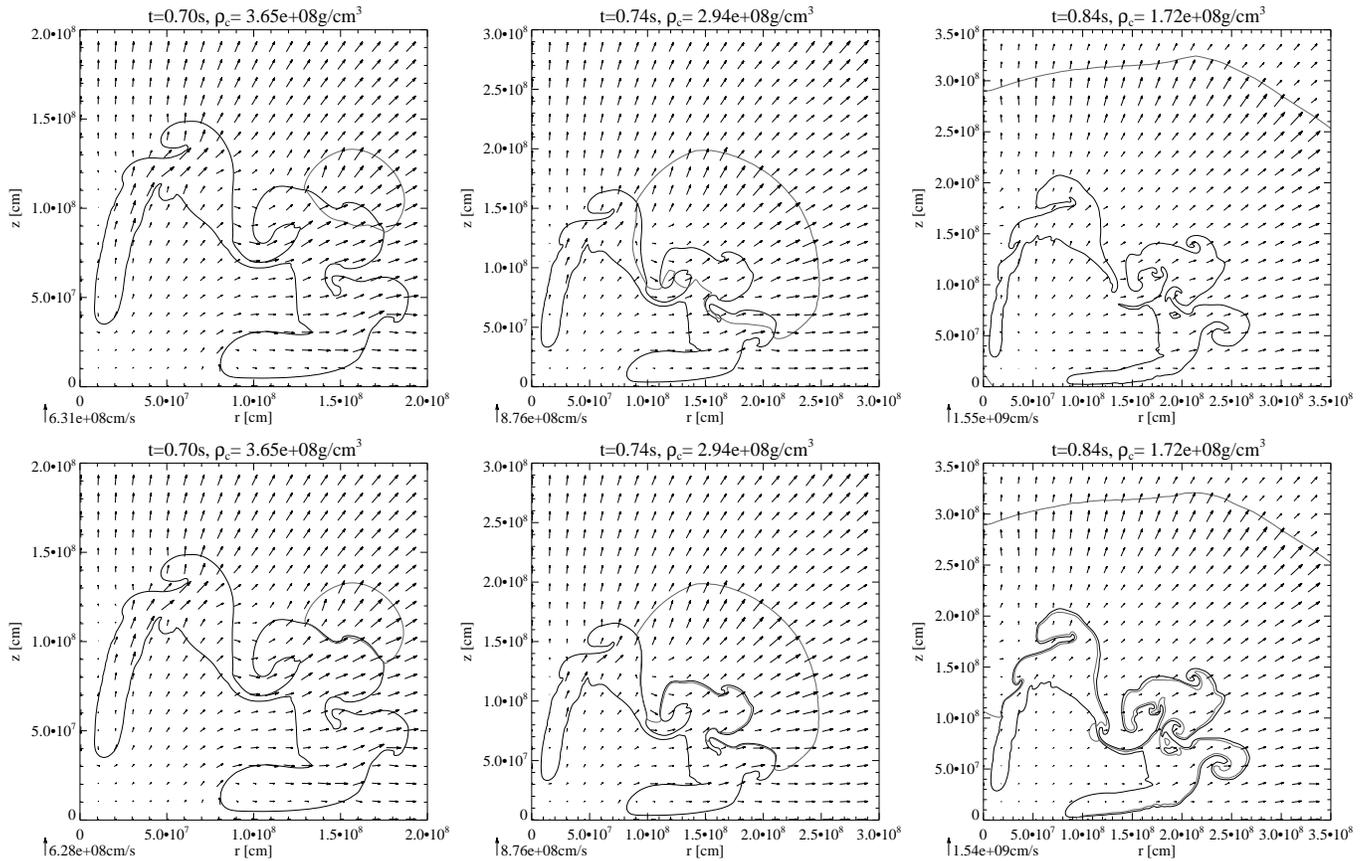}
}
\caption{Temporal evolution of the front geometry and velocity field
for the Models B1a (top row) and B1b (bottom row). The snapshots were taken at
equal times. The convoluted front corresponds to the initial deflagration, and
the smooth curve that grows with time depicts the detonation front.}
\label{fig:b1}
\end{figure*}

The time evolution of the front geometry and flow field in the
non-central one-bubble model B1 is shown in Fig.\ \ref{fig:b1} where
we compare three snapshots of the explosion scenarios a and b at
$t=0.7\,\mathrm{s}$, $0.74\,\mathrm{s}$ and $0.84\,\mathrm{s}$.
As an additional measure for the expansion of the star, the central
density is noted in every plot. By construction the deflagration phase 
is identical for both Cases a and b. 

After the DDT which takes place at $t = 0.68\,\mathrm{s}$ 
at a distance $r\approx 1.5 \times 10^8\,\mathrm{cm}$, the differences
between Models B1a and B1b become visible. The detonation in
Model B1b cannot cross the bubble
of burnt material until the latter breaks apart at its thinnest point at $t
\approx 0.84$ s  (lower right plot of Fig.\ \ref{fig:b1}). Notice that the
breaking is only
a numerical artifact since burnt regions cannot disconnect in the absence of
flame quenching; more realistic, i.e.\ highly resolved and three-dimensional,
simulations are needed to study the effects of ash clumps on predictions
of the delayed detonation scenario.

For completeness, the global properties of Models ZIa to B5a are listed in
Table \ref{global_tab}. In all of these models, no isolated regions of fuel
developed at late times, hence the Case b simulations were nearly identical to
Case a in terms of global quantities.

\begin{table}
\centering
\caption{Global characteristics of Case "a" models (detonation re-ignited after
  passing through clumps of ashes).
\label{global_tab}}
\begin{tabular}{|c||c|c|c|c|} \hline 
model & $\mathrm{E_{nuc}}$ & $\mathrm{E_{kin}}$ &
$\mathrm{M_{Mg}}$  & $\mathrm{M_{Ni}}$ \\ 
&$\mathrm{ [10^{51}\,erg]}$ & $\mathrm{ [10^{51}\,erg]}$ & $\mathrm{
  [M_{\odot}]}$  & $\mathrm{ [M_{\odot}]}$ \\ \hline \hline
Z1a & 1.53 & 0.90 & 0.66 & 0.61  \\ \hline
Z3a & 1.44 & 0.87 & 0.58 & 0.65  \\ \hline
B1a & 1.75 & 1.19 & 0.87 & 0.42  \\ \hline
B5a & 1.51 & 0.95 & 0.68 & 0.53  \\ \hline
\end{tabular}
\end{table}

As can be seen from the energies and nickel amounts, all four models 
produce healthy explosions with high total energies. 
The off-center ignition Scenarios B5 and especially B1 are more
powerful than the central explosion Models Z1 and Z3. One reason for
this is that an off-center deflagration phase leaves large 
amounts of unburnt material at high densities near the center of the
white dwarf, whereas a centrally ignited flame causes most material to 
expand significantly before the detonation sets in. In addition, the
transition from deflagration to detonation  
took place earlier ($0.6\,\mathrm{s}$ -$0.7\,\mathrm{s}$) in the 
off-center than in the central ($\sim0.85\,\mathrm{s}$) models.
 
\subsection{Conclusions}
\label{sec:conclusions}

In order to simulate delayed detonations in multidimensional explosion
models of type Ia supernovae, we modified and tested
a front algorithm previously developed to model subsonic deflagrations. The
test calculations with detonation fronts presented in 
this paper show that, in spite of a significantly higher burning
speed and density jump across the front, the inaccuracies in
reproducing the correct front velocity are the same as those 
for deflagrations ($\lesssim 10$ \%). In
particular, the deviations were shown to be 
independent of the flow velocity in the grid frame and therefore easily
correctable by renormalizing the prescribed front velocity. 

In a first exploratory set of two-dimensional explosion simulations
whose deflagration phases were identical to those in \citet{RHN99},
we used two independent implementations of the front model, one describing the
turbulent flame front in the deflagration stage and the 
other accounting for the detonation front after an assumed
deflagration-detonation-transition (DDT). The time and location of the
DDT were determined by tracking the ratio of Gibson length
and thermal flame thickness, and triggering the detonation where this parameter
dropped below unity \citep[following][]{NW97,NK97a}. This procedure
gave rise to DDTs at radii roughly $1.5 \times 10^8$ cm off-center
and times of  $0.6 \dots 0.8$ s after ignition of the
deflagration. Off-center deflagration models generically resulted in
higher energy release as a consequence of the earlier DDT and
correspondingly higher core fuel density during the detonation phase. 

More realistic three-dimensional simulations are planned for future work. We
expect a noticeable difference as a result of the changed flame front
topology. However, the majority of scales of unburned clumps won't be
resolved in these simulations and will have to be modeled on subgrid scales. 

We emphasize the importance of using a tracking/capturing scheme for the
detonation, as well as for the deflagration, as opposed to the traditional
approach to model detonations as unresolved supersonic burning
fronts. In addition to being able to use accurate tabulated detonation
velocities as a function of the unburned fuel density, it allows full
control over the interaction of the detonation with regions of burned
material. The viability of the delayed detonation scenario hinges on
the capability of the detonation to burn the remaining C+O in the
core, and hence on its ability to cross the foam of fuel and ashes in
the Rayleigh-Taylor mixing region. These questions will be investigated
in a forthcoming publication.

\acknowledgements
We are grateful for discussions with Christian Klingenberg, Andreas Maier, Jan
Pfannes, Martin Reinecke, Fritz Röpke, and Wolfram Schmidt. The research of
JCN was supported by the Alfried Krupp Prize for Young 
University Teachers of the Alfried Krupp von Bohlen und Halbach Foundation.

\bibliographystyle{aa}
\bibliography{snrefs}

\end{document}